\documentclass[prl,twocolumn,showpacs,superscriptaddress]{revtex4-1}
\usepackage{graphicx}
\usepackage{units}
\usepackage{multirow}
\usepackage{bigstrut}
\usepackage{overpic}
\usepackage{array}
\usepackage{color}
\usepackage{amsmath}
\usepackage{bm}
\usepackage{times}
\RequirePackage{lineno}
\begin{document}

\modulolinenumbers[2]

\setlength{\oddsidemargin}{-0.5cm} \addtolength{\topmargin}{15mm}

\title{\boldmath First Measurement of the Form Factors in $D^+_{s}\rightarrow K^0 e^+\nu_e$ and $D^+_{s}\rightarrow K^{*0} e^+\nu_e$ Decays}

\author{
  \small
   M.~Ablikim$^{1}$, M.~N.~Achasov$^{10,d}$, S.~Ahmed$^{15}$, M.~Albrecht$^{4}$, M.~Alekseev$^{55A,55C}$, A.~Amoroso$^{55A,55C}$, F.~F.~An$^{1}$, Q.~An$^{52,42}$, Y.~Bai$^{41}$, O.~Bakina$^{27}$
% ...
   \\
   \vspace{0.2cm}
   (BESIII Collaboration)\\
   \vspace{0.2cm} {\it
   $^{1}$ Institute of High Energy Physics, Beijing 100049, People's Republic of China\\
% ...
$^{4}$ Bochum Ruhr-University, D-44780 Bochum, Germany\\
% ...
$^{10}$ G.I. Budker Institute of Nuclear Physics SB RAS (BINP), Novosibirsk 630090, Russia\\
% ...
$^{15}$ Helmholtz Institute Mainz, Johann-Joachim-Becher-Weg 45, D-55099 Mainz, Germany\\
% ...
$^{27}$ Joint Institute for Nuclear Research, 141980 Dubna, Moscow region, Russia\\
% ...
$^{41}$ Southeast University, Nanjing 211100, People's Republic of China\\
   $^{42}$ State Key Laboratory of Particle Detection and Electronics, Beijing 100049, Hefei 230026, People's Republic of China\\
% ...
$^{52}$ University of Science and Technology of China, Hefei 230026, People's Republic of China\\
% ...
$^{55}$ (A)University of Turin, I-10125, Turin, Italy; (B)University of Eastern Piedmont, I-15121, Alessandria, Italy; (C)INFN, I-10125, Turin, Italy\\
% ...
$^{d}$ Also at the Novosibirsk State University, Novosibirsk, 630090, Russia\\
% ...
   \vspace{0.4cm}
}
}

\begin{abstract}
We report on new measurements of Cabibbo-suppressed semileptonic
$D_s^+$ decays using $3.19~\mathrm{fb}^{-1}$ of $e^+e^-$
annihilation data sample collected at a center-of-mass energy of
4.178~GeV with the BESIII detector at the BEPCII collider.  Our
results include branching fractions $\mathcal B({D^+_s\rightarrow
K^0 e^+\nu_{e}})=(3.25\pm0.38({\rm stat.})\pm0.16({\rm
syst.}))\times10^{-3}$ and $\mathcal B({D^+_s\rightarrow K^{*0}
e^+\nu_{e}})=(2.37\pm0.26({\rm stat.})\pm0.20({\rm
syst.}))\times10^{-3}$ which are much improved relative to
previous measurements, and the first measurements of the hadronic
form-factor parameters for these decays.   For $D^+_s\rightarrow
K^0 e^+\nu_{e}$, we obtain $f_+(0)=0.720\pm0.084({\rm
stat.})\pm0.013({\rm syst.})$, and for $D^+_s\rightarrow K^{*0}
e^+\nu_{e}$, we find form-factor ratios
$r_V=V(0)/A_1(0)=1.67\pm0.34({\rm stat.})\pm0.16({\rm syst.})$ and
$r_2=A_2(0)/A_1(0)=0.77\pm0.28({\rm stat.})\pm0.07({\rm
syst.})$.
\end{abstract}


\maketitle

The study of $D^+_s$ semileptonic\,(SL) decays provides valuable
information about weak and strong interactions in mesons composed of
heavy quarks.  (Throughout this Letter, charge-conjugate modes are
implied unless explicitly noted.) Measurement of the total SL decay
width of the $D^+_s$, and comparison with that of the $D$ mesons,
can help elucidate the role of nonperturbative effects in
heavy-meson decays~\cite{plb515,prd80_052007}. The
Cabibbo-suppressed~(CS) SL decays, including the branching fractions
(BFs) for $D^+_{s}\rightarrow K^0 e^+\nu_e$ and $D^+_{s}\rightarrow
K^{*0} e^+\nu_e$~\cite{pdg}, are especially poorly measured.
Detailed investigations of the dynamics of these decays allow
measurements of SL decay partial widths, which depend on the
hadronic form factors (FFs) describing the interaction between the
final-state quarks.  Measurements of these FFs provide experimental
tests of theoretical predictions of lattice QCD~(LQCD).
Reference~\cite{lattice} predicts that the FFs have minimal
dependence on the spectator-quark mass, with values for
$D^+_s\rightarrow K^{0}\ell^+\nu_{\ell}$ and $D^+\rightarrow
\pi^0\ell^+\nu_{\ell}$ differing by less than 5\%.  
Experimental verification of this predicted instance would be a significant success for LQCD.
A complementary LQCD test is provided by comparing measured
and predicted FF parameters for $D^+_s\rightarrow
K^{*0}\ell^+\nu_{\ell}$ and $D^+ \rightarrow
\rho^0\ell^+\nu_{\ell}$. The combination of these measurements has
the potential to verify LQCD FF predictions for SL charm decays to
both pseudoscalar and vector mesons, useful for further applying the
LQCD to SL $B$ decays for precise determination of
Cabibbo-Kobayashi-Maskawa\,~(CKM)
parameters~\cite{lattice,EPJC74_2981,prd85_114502}.

In this Letter, we report on improved measurements of the absolute BFs and first measurements of the FFs for
the decays $D_s^+\rightarrow K^0e^+\nu_e$ and $D_s^+\rightarrow K^{*0}e^+\nu_e$.  Our measurements
have been made with $3.19~\mathrm{fb}^{-1}$ $e^+e^-$ annihilation data recorded with the BESIII detector at the BEPCII collider.  The
center-of-mass energy for our data is $\sqrt{s}=4.178$ GeV. The cross section is $\sim 1$~nb
for the production of $D_s^{*+}D_s^- + c.c.$ at this energy. Our data sample is the largest collected by any
experiment for $D^+_s$ studies in the clean near-threshold environment.

Details about the BESIII detector design and performance are provided in Ref.~\cite{Ablikim:2009aa}.
A {\sc geant4}-based~\cite{geant4} Monte Carlo (MC) simulation package,
which includes the geometric description of the detector and the
detector response, is used to determine signal detection efficiencies and
to estimate potential backgrounds. Signal MC samples of $e^+e^-\rightarrow D_s^{*+}D_s^-$ with a
$D^+_s$ meson decaying to $K^{(*)0}e^+\nu_e$ together with a
$D^-_s$ decaying to the studied decay modes used for this analysis are generated
with {\sc conexc}~\cite{conexc} using
{\sc evtgen}~\cite{nima462_152}, 
with the inclusion of initial-state radiation (ISR) effects up to second order correction~\cite{conexc,SJNP41_466}. The final-state radiation
(FSR) effects are simulated via the {\sc PHOTOS} package~\cite{plb303_163}. The interference effects between ISR and FSR are ignored~\cite{prd63_113009}.
The simulation of the SL decay
$D_s^+\rightarrow K^{*(0)}e^+\nu_e$ is matched with the FFs measured in this work.
To study the backgrounds, inclusive MC samples consisting of open-charm states, radiative
return to $J/\psi$ and $\psi(2S)$ and continuum processes of $q\bar{q}$ ($q=u,d,s$), along with Bhabha scattering, $\mu^+\mu^-$, $\tau^+\tau^-$ and $\gamma\gamma$ events are generated.
All known decay modes of open-charm and $\psi$ states are
simulated as specified by the Particle Data Group (PDG)~\cite{pdg15}, while the
remaining unknown decays are modeled with {\sc lundcharm}~\cite{lundcharm}.


As described above, $D_s^+$ mesons are produced at $\sqrt{s}=4.178$~GeV predominantly
through $D_s^{*+}D_s^-$~\cite{prd80_072001}, with 94\% of the $D^{*+}_s$ decaying to
$\gamma D_s^+$.   The first step of our analysis is to select ``single-tag" (ST) events with a fully
reconstructed $D_{s}^-$ candidate. The $D_s^-$ hadronic decay tag modes that are used for this analysis are listed in Table~\ref{tab:numST}.
In this ST sample, we select the SL decay $D_s^+\rightarrow K^{(*)0}e^+\nu_e$
plus an isolated photon consistent with being from the $D_s^* \rightarrow \gamma D_s$ transition.  The
selected events are referred to as the double-tag (DT) sample.  For a specific tag mode $i$, the ST and DT
event yields can be expressed as
$$N^{i}_{\rm ST}=2N_{D_sD_s^{*}}\mathcal{B}^i_{\rm ST}\epsilon^i_{\rm
ST}~~{\rm and}~~N^{i}_{\rm DT}=2N_{D_sD_s^{*}}\mathcal{B}^i_{\rm
ST}\mathcal{B}^i_{\rm SL}\epsilon^i_{\rm DT},$$
where $N_{D_sD_s^{*}}$ is the number of $D_sD_s^{*}$ pairs; $\mathcal{B}^i_{\rm ST}$ and
$\mathcal{B}^i_{\rm SL}$ are the BFs of the $D_s^-$ tag mode and the $D_s^+$ SL decay
mode, respectively; $\epsilon^i_{\rm ST}$ is the efficiency for finding the tag candidate; and
$\epsilon^i_{\rm DT}$ is the efficiency for simultaneously finding the tag $D^-_s$ and the SL decay.
The DT efficiency $\epsilon^i_{\rm DT}$ includes the BF for $D^{*+}_s\rightarrow \gamma D^+_s$.
The BF for the SL decay is given by

\begin{equation}
\mathcal{B}_{\rm SL}=\frac{N_{\rm DT}}{\sum N^{i}_{\rm
ST}\times\epsilon^i_{\rm DT}/\epsilon^i_{\rm ST}}=\frac{N_{\rm DT}}{N_{\rm ST}\times\epsilon_{\rm SL}}, \label{eq:branch}
\end{equation}
where $N_{\rm DT}$ is the total yield of DT events, $N_{\rm ST}$ is the total ST yield, and
$\epsilon_{\rm SL}=\frac{\sum N^{i}_{\rm ST}\times\epsilon^i_{\rm DT}/\epsilon^i_{\rm ST}}{\sum N^{i}_{\rm ST}}$
is the average efficiency for finding SL decay weighted by the measured yields of tag modes in data.

Selection criteria for $\gamma$, $\pi^\pm$ and $K^\pm$ are the same as those used in
Ref.~\cite{prl118_112001}. $\pi^0(\eta)$ candidate is reconstructed from the $\gamma\gamma$ combination with invariant mass within $(0.115,~0.150)\,[(0.50,~0.57)]$~GeV$/c^2$. To improve
the momentum resolution, a kinematic fit is performed to constrain  $\gamma\gamma$ invariant mass to the
nominal $\pi^0(\eta)$ mass~\cite{pdg} with $\chi^2<20$. The
fitted $\pi^0\,(\eta)$ momenta are used for further analysis.
$K^0_S$ mesons are reconstructed from two oppositely charged tracks with its invariant mass within $(0.485,~0.510)$~GeV$/c^2$. 
A vertex constriant is applied to improve $K^0_S$ signal significance as in Ref.~\cite{1811.11349}.
We select $\rho^{-} \rightarrow \pi^-\pi^{0}$ by requiring the invariant
mass $M_{\pi^-\pi^{0}}$ to be within $(0.626,~0.924)$~GeV/$c^2$~\cite{pdg}.  The decay modes
$\eta^{\prime}\rightarrow \pi^+\pi^-\eta$ and $\eta^{\prime}\rightarrow \gamma\pi^+\pi^-$ are used to select
$\eta^{\prime}$ mesons, with the invariant masses of the $\pi^+\pi^-\eta$ and $\gamma\pi^+\pi^-$
required to be within $(0.940,~0.976)$~GeV$/c^2$ and $(0.940,~0.970)$~GeV$/c^2$, respectively.
Additionally, to suppress backgrounds from $D^*$ decays, the momenta of the photons
from $\eta^{\prime}\rightarrow \gamma\pi^+\pi^-$ and all pions are required to be
greater than 0.1 GeV/$c$.

\begin{table}[tp!]
\caption{ $M_{D_s^-}$ windows and ST yields in data. }
\begin{center}
\begin{tabular}
{lcc} \hline ST mode~~~ & $M_{D_s^-}$ (GeV/$c^2$) &  $N^i_{\rm ST}$   \\
\hline
$K_S^0K^-$                        &  (1.945, 1.990) & ~~$25858\pm217$        \\
$K^+K^-\pi^-$                     &  (1.945, 1.990) & $130666\pm575$         \\
$K_S^0K^-\pi^0$                   &  (1.940, 1.990) & ~~$10807\pm398$        \\
$K_S^0K_S^0\pi^-$                 &  (1.945, 1.990) & ~~~~$3810\pm131$       \\
$K^+K^-\pi^-\pi^0$                &  (1.940, 1.990) & ~~$35091\pm702$        \\
$K_S^0K^-\pi^+\pi^-$              &  (1.945, 1.990) & ~~~~$7722\pm235$       \\
$K_S^0K^+\pi^-\pi^-$              &  (1.945, 1.990) & ~~$14802\pm259$        \\
$\pi^+\pi^-\pi^-$                 &  (1.945, 1.990) & ~~$36258\pm832$        \\
$\pi^-\eta$                       &  (1.940, 1.990) & ~~$17535\pm400$        \\
$\rho^-\eta$                      &  (1.940, 1.990) & ~~$30114\pm886$        \\
$\pi^-\eta^{\prime}(\eta^{\prime}\rightarrow\pi^+\pi^-\eta)$   &  (1.940, 1.990) & ~~~~$7704\pm152$  \\
$\rho^-\eta^{\prime}(\eta^{\prime}\rightarrow\pi^+\pi^-\eta)$  &  (1.940, 1.990) & ~~~~$3039\pm226$  \\
$\pi^-\eta^{\prime}(\eta^{\prime}\rightarrow\gamma\pi^+\pi^-)$ &  (1.940, 1.990) & ~~$17919\pm481$   \\
\hline
\end{tabular}
\label{tab:numST}
\end{center}
\end{table}

\begin{figure}[tp!]
\begin{center}
\includegraphics[width=\linewidth]{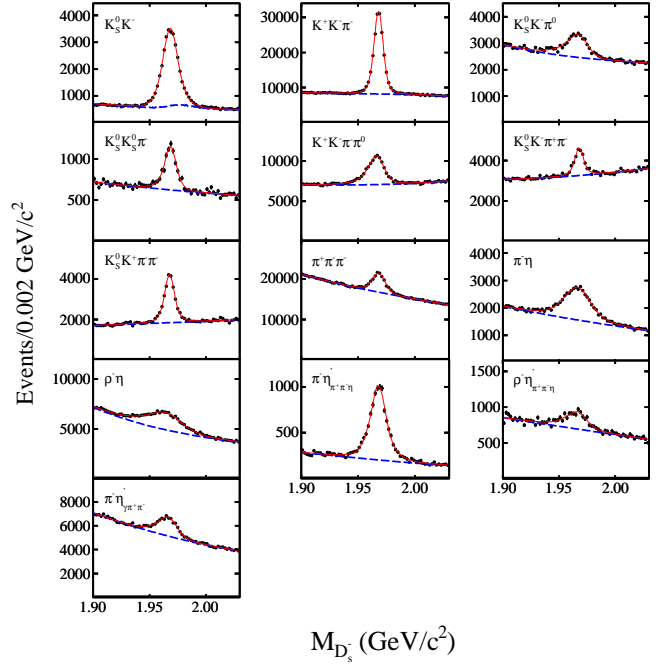}
\caption{(Color online)
Fits to $M_{D_s^-}$ distributions for the thirteen tag modes.
Points with error bars are data, blue dashed curves are the fitted backgrounds and
red solid curves are the total fits.}
\label{fig:tag_mds}
\end{center}
\end{figure}

For all events passing the ST selection criteria, we calculate the recoil mass against the tag with the following formula:

$$M_{\rm rec}=\sqrt{(\sqrt{s}-\sqrt{|\vec{p}_{D_s^-}|^2+m^2_{D_s^-}})^2-|\vec{p}_{D_s^-}|^2},$$
where $m_{D^-_s}$ and $\vec{p}_{D_s^-}$ are the
known mass~\cite{pdg} and measured momentum of the tag $D_s^-$.  We define
$\Delta M\equiv M_{\rm rec}-m_{D_s^{*+}}$,
where $m_{D_s^{*+}}$ is the nominal $D_s^{*+}$ mass~\cite{pdg}.  Events within
$-0.060<\Delta M<0.065$~GeV/$c^2$ are accepted as $D_s^{*+}D_s^-$ candidates.
To extract the mode-by-mode ST yields, we perform unbinned maximum likelihood fits to the distributions
of the $D^-_s$ invariant mass $M_{D_s^-}$, as shown in Fig.~\ref{fig:tag_mds}.
Signals are modeled with the MC-simulated signal shape convoluted with Gaussians to account for
the resolution differences between data and MC simulation, while the
combinatorial backgrounds are parametrized with second- or third-order polynomial
functions.
Because of the misidentification of $\pi^-$ as $K^-$,
the backgrounds from $D^-\rightarrow K^0_S\pi^-$ form a broad peak near the $D_s^-$ nominal mass for $D_s^-\rightarrow K^0_SK^-$.
In the fit, the shape of this background is described by using the MC simulation and its size is set as a free parameter.  For each tag mode, the ST yield is obtained by integrating the signal function over the $D^-_s$ mass signal region specified in the second column of Table~\ref{tab:numST}, which also
includes the ST yields for all tag modes.  The total reconstructed ST yield in our data sample is $N_{\rm ST}=341,325\pm1,764$.

\begin{linenumbers}
In signal events, the system recoiling against the $D_s^-$ tag consists of the SL decay
$D_s^+\rightarrow K^0e^+\nu_e$ or $D_s^+\rightarrow K^{*0}e^+\nu_e$.
We select these from the additional tracks accompanying the tag, that is a $K^0\rightarrow K^0_S\rightarrow \pi^+\pi^-$ with the ST criteria already described,
and $K^{*0}\rightarrow K^+\pi^-$, therefore requiring that there be exactly three tracks in the event and with the invariant mass $M_{K^+\pi^-}$ required to be within
$(0.801,~0.991)$~GeV/$c^2$.
Detection and reconstruction of the positron follow the procedures in Refs.~\cite{bes3electronSL,prl115_221805}.
Backgrounds from $D_s^+\rightarrow K^0\pi^+$ reconstructed as $D_s^+\rightarrow K^{0}e^+\nu_e$ and $D_s^+\rightarrow K^+\pi^+\pi^-$ reconstructed as $D_s^+\rightarrow K^{*0}e^+\nu_e$ are rejected by requiring the $K^0e^+$ or $K^{*0}e^+$ invariant mass to be less than 1.78~GeV/$c^2$.
Backgrounds associated with fake photons are suppressed by requiring $E_{\gamma \rm{max}}$,
the largest energy of any unused photon, to be less than 0.20~GeV.

To identify a photon produced directly from $D_s^{*\pm}$, we perform two kinematic fits for each $\gamma$ candidate, one assuming that the $\gamma$ combines
with the tag to form a $D_s^{*-}$ and the other assuming that the SL decay comes from a $D_s^{*+}$ parent.  We require
the $D^{\mp}_sD_s^{*\pm}$ pair to conserve energy and momentum in the center-of-mass frame, and the
$D^{\pm}_s$ candidates are constrained to the known mass.  The neutrino is treated as a missing particle.  When we assume the tag to be
the daughter of a $D_s^{*-}$, we constrain the mass of the photon plus tag candidate to be consistent with the expected
$D_s^{*-}$ mass; otherwise we constrain the mass of the photon plus SL decay to be consistent with the $D_s^{*+}$mass.
Finally, we select the photon and hypothesis with the smallest kinematic fit $\chi^2$.

We obtain information about the undetected neutrino with the missing-mass squared (${\rm MM}^2$) of the event, calculated from
the energies and momenta of the tag ($E_{D_s^-}$, $\vec{p}_{D_s^-}$), the transition photon ($E_{\gamma}$,
$\vec{p}_{\gamma}$), and the detected SL decay products ($E_{\rm SL}=E_{K^{(*)0}}+E_{e^+}$,
$\vec{p}_{\rm SL}=\vec{p}_{K^{(*)0}}+\vec{p}_{e^+}$) as follows:
\end{linenumbers}
$${\rm MM}^2=(\sqrt{s}-E_{D_s^-}-E_{\gamma}-E_{\rm SL})^2-(|\vec{p}_{D_s^-}+\vec{p}_{\gamma}+\vec{p}_{\rm SL}|)^2.$$
\begin{linenumbers}

\begin{figure}[tp!]
\begin{center}
   \begin{minipage}[t]{8.8cm}
   \includegraphics[width=\linewidth]{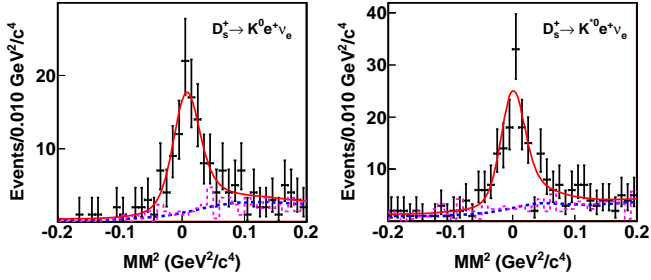}
   \end{minipage}
   \caption{(Color online)~
   Fits to ${\rm MM}^2$ distributions of SL candidate events.
   Dots with error bars are data, dot-dashed lines (blue) are the fitted backgrounds
   and solid curves (red) are the total fits. The long-dashed lines (pink) show the backgrounds from the $M_{D_s^-}$ sidebands. }
\label{fig:m2miss}
\end{center}
\end{figure}

\end{linenumbers}
Figure~\ref{fig:m2miss} shows the ${\rm MM}^2$ distributions of the accepted candidate events for $D_s^+\rightarrow K^{0}e^+\nu_e$
and $D_s^+\rightarrow K^{*0}e^+\nu_e$ in data.
The signal DT yield $N_{\rm DT}$ is obtained by performing an unbinned maximum likelihood fit to ${\rm MM}^2$.
In the fit, the signal is described with a MC-derived signal shape convolved with a Gaussian,
and the background is described by a shape obtained from the inclusive MC sample, in which no peaking backgrounds are observed.
We obtain $117.2\pm13.9$ and $155.0\pm17.2$ events for $D_s^+\rightarrow K^0e^+\nu_e$ and $D_s^+\rightarrow
K^{*0}e^+\nu_e$, respectively, where the uncertainties are statistical only.
No peaking backgrounds are observed in $K^{(*)0}$ mass sideband.

The BFs of $D_s^+\rightarrow K^{0}e^+\nu_e$ and $D_s^+\rightarrow K^{*0}e^+\nu_e$ are determined by Eq.~(\ref{eq:branch}), where the detection efficiencies $\varepsilon_{\rm SL}$ are estimated to be $(10.57\pm0.04)\%$ and $(19.15\pm0.06)\%$ for
$D_s^+\rightarrow K^0e^+\nu_e$ and $D_s^+\rightarrow K^{*0}e^+\nu_e$, respectively.  (These efficiencies include the
BFs for $K^0\rightarrow \pi^+\pi^-$ and $K^{*0}\rightarrow K^+\pi^-$.)  
Finally, we obtain $\mathcal B({D^+_s\rightarrow K^0 e^+\nu_{e}})=(3.25\pm0.38)\times10^{-3}$ and $\mathcal
B({D^+_s\rightarrow K^{*0} e^+\nu_{e}})=(2.37\pm0.26)\times10^{-3}$,
where the uncertainties are statistical only.

With the DT technique, the BF measurements are insensitive to the systematic uncertainties of the ST selection.
The uncertainties of the $e^+$ tracking and PID efficiencies have all been determined to be 1.0\%~\cite{prl115_221805}, while the uncertainty
of the $K^{(*)0}$ reconstruction is 1.5\,(2.3)\%.
The uncertainty associated with the ${\rm MM}^2$ fit is estimated to be 3.5\,(3.8)\% by varying the fitting ranges and
the signal and background shapes.
The uncertainty due to the selection of the $\gamma$ is estimated to be 2.0\% based on selecting the best photon
candidate in a control sample of $e^+e^-\rightarrow D_s^{+*}D_s^-$ events with two hadronic tags,
$D_s^+\rightarrow K_S^0K^+$ and $D_s^-\rightarrow K^+K^-\pi^-$.
The uncertainties due to the $E_{\gamma \rm{max}}$ and $M_{K^{(*)0}e^+}$ requirements
are estimated to be 1.7\,(1.7)\% and 0.7\,(0.9)\% by comparing the nominal BF with that measured with alternative requirements.
The uncertainty due to the MC signal modeling is estimated to be 0.9\,(1.8)\% by varying
the input FF parameters by $\pm 1\sigma$ as determined in this work.
We also consider the systematic uncertainties of $N_{\rm ST}$ (0.5\%), evaluated by using
alternative signal shapes when fitting the $M_{D_s^-}$ spectra, and of the MC statistics (0.4\,[0.3]\%).
The uncertainty due to different tag dependences between data and MC simulation is estimated to be 0.8\,(0.3)\%.
Additionally, for $D_s^+\rightarrow K^{*0}e^+\nu_e$ decay,
the systematic uncertainty for the possible $\mathcal{S}$\verb|-|wave component in the $K\pi$ system is estimated to be $6.0\%$ according to Refs.~\cite{prd83_072001,prd94_032001}.
Adding these contributions in quadrature gives total systematic uncertainties of 5.1\% and 8.3\% for
$\mathcal B({D^+_s\rightarrow K^{0} e^+\nu_{e}})$ and $\mathcal B({D^+_s\rightarrow K^{*0} e^+\nu_{e}})$, respectively.

\begin{figure}[tp!]
\begin{center}
   \flushleft
   \begin{minipage}[t]{8.8cm}
   \includegraphics[width=\linewidth]{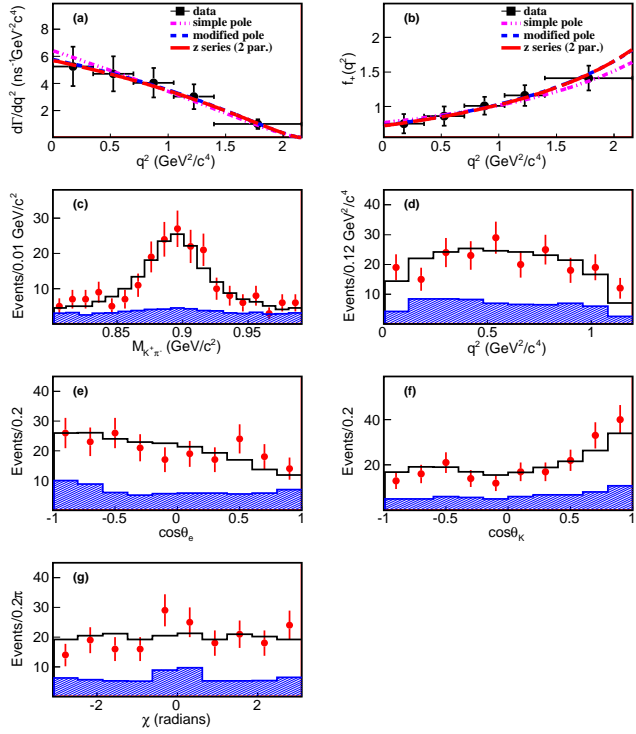}
   \end{minipage}
   \caption{(Color online)
   (a) Fits to the differential decay rates and (b) projections onto $f^{K}_+(q^2)$ for $D_s^+\rightarrow K^0e^+\nu_e$.
   Projections onto (c) $M_{K^+\pi^-}$, (d) $q^2$, (e) $\cos\theta_e$, (f) $\cos\theta_K$, and (g) $\chi$ for $D_s^+\rightarrow K^{*0}e^+\nu_e$.
   Dots with error bars are data. Curves in (a,\,b) give the best fits with different FF parametrizations.
   Solid and shadowed histograms in (c,\,d,\,e,\,f,\,g) are the MC-simulated signal plus background
   and the MC-simulated background.  }
\label{fig:formfactor}
\end{center}
\end{figure}

The $D^+_s\rightarrow K^0 e^+\nu_{e}$ differential decay width
with respect to the mass squared ($q^2$) of the $e^+\nu_{e}$ system is expressed as~\cite{prd78_054002}:
\begin{equation}
\frac{d\Gamma(D_s^+\rightarrow K^0e^+\nu_e)}{dq^2}=\frac{G^2_F|V_{cd}|^2}{24\pi^3}p^3_{K^0}|f^{K}_+(q^2)|^2.
\label{eq:dgammadq2_ksev}
\end{equation}
In this equation $p_{K^0}$ is the $K^0$ momentum in the rest frame of the $D_s^+$, $G_F$ is the Fermi
constant~\cite{pdg}, $|V_{cd}|$ is the CKM matrix element, and $f_+^K(q^2)$ is the hadronic FF.
To extract the FF parameters, we fit to the differential decay rates $\Delta\Gamma_i$ measured in the $q^2$ bins of [0.00, 0.35), [0.35, 0.70), [0.70, 1.05), [1.05, 1.40) and [1.40, 2.16)~GeV/$c^2$ by using the three theoretical parametrizations in
Table~\ref{tab:sum_formfactor}.  A least-$\chi^2$ fit is performed accounting for correlations among $q^2$ bins.
We fix the pole mass $m_{\rm pole}$ at the $D^{*+}$ nominal mass~\cite{pdg}.
The fits to the differential decay rate and projections of the fits onto $f_+(q^2)$ for $D_s^+\rightarrow K^0e^+\nu_e$
are shown in Figs.~\ref{fig:formfactor}(a) and (b), and the FF fit results are summarized in the third column of
Table~\ref{tab:sum_formfactor}.
The systematic uncertainties in the extracted parameters are estimated as in Ref.~\cite{prd96_012002}.
These include the same systematic effects as the BF measurements, along with the $D^+_s$-lifetime uncertainty.
Using $|V_{cd}|=0.22492\pm0.00050$~\cite{pdg},
we obtain $f^{K}_+(0)$ as shown in the last column of Table~\ref{tab:sum_formfactor}.

\begin{table}
\begin{center}
\caption{FF results from fits to $D^+_s\rightarrow K^0 e^+\nu_{e}$, where the first errors are statistical and the second systematic.}
\resizebox{!}{0.85cm}{
\begin{tabular}
{l|c|c}
\hline\hline
Parametrizations                            &  $f^{K}_+(0)|V_{cd}|$                   & $f^{K}_+(0)$             \\
\hline
Simple pole~\cite{plb478_417}                & ~~$0.172\pm0.010\pm0.001$               &  $0.765\pm0.044\pm0.004$ \\
Modified pole~\cite{plb478_417}              & ~~$0.163\pm0.017\pm0.003$               &  $0.725\pm0.076\pm0.013$ \\
z series (two par.)~\cite{plb633_61}           & ~~$0.162\pm0.019\pm0.003$               &  $0.720\pm0.084\pm0.013$ \\
\hline \hline
\end{tabular}
}
\label{tab:sum_formfactor}
\end{center}
\end{table}
\normalsize

The differential decay rate of $D^+_s\rightarrow K^{*0} e^+\nu_{e}$ depends on five variables:
$K\pi$ mass squared ($m_{K\pi}^2$), $e^+\nu_e$ mass squared ($q^2$), the angle between the
$K^+$ and $D_s^+$ momenta in the $K\pi$ rest frame ($\theta_K$),
the angle between the $\nu_e$ and $D^+_s$ momenta in the $e^+\nu_e$ system ($\theta_e$), and the
acoplanarity angle between the $K\pi$ and $e^+\nu_e$ decay planes ($\chi$).
The differential decay rate can be expressed in terms of three helicity amplitudes~\cite{prl110_131802,prd92_071101}:
$H_{\pm}(q^2)=(M_{D^+_s}+m_{K\pi})A_1(q^2)\mp\frac{2M_{D^+_s} P_{K\pi}}{M_{D^+_s}+M_{K\pi}}V(q^2)$ and
$H_0(q^2)=\frac{1}{2m_{K\pi}q}[(M_{D^+_s}^2-m_{K\pi}^2-q^2)(M_{D^+_s}+m_{K\pi})A_1(q^2)-\frac{4M_{D^+_s}^2p^2_{K\pi}}{M_{D^+_s}+M_{K\pi}}A_2(q^2)]$,
where $p_{K\pi}$ is the momentum of the $K\pi$ system in the rest frame of the $D_s^+$,
and $V(q^2)$ and $A_{1/2}(q^2)$ are the vector and axial FFs, respectively.
Because $A_1(q^2)$ is common to all three helicity amplitudes,
it is natural to define the FF ratios $r_V=V(0)/A_1(0)$ and $r_2=A_2(0)/A_1(0)$.
The $A_{1/2}(q^2)$ and $V(q^2)$ are assumed to have simple pole forms,
$A_{1/2}(q^2)=A_{1/2}(0)/(1-q^2/M^2_A)$ and $V(q^2)=V(0)/(1-q^2/M^2_V)$, with pole masses
$M_V=M_{D^*(1^-)}=2.01$~GeV/$c^2$ and $M_A=M_{D^*(1^+)}=2.42$~GeV/$c^2$~\cite{pdg}.

We perform a five-dimensional maximum likelihood fit in the space of $M^2_{K^+\pi^-}$, $q^2$, $\cos\theta_e$, $\cos\theta_K$, and $\chi$ for the $D^+_s\rightarrow K^{*0}e^+\nu_e$ events within $-0.15<{\rm MM}^2<0.15$ GeV$^2$/$c^4$ in a similar manner as Refs.~\cite{prl110_131802,prd92_071101}.
We ignored the possible $\mathcal{S}$\verb|-|wave component in $K\pi$ system due to limited statistics.
The projections of the fit onto $M^2_{K^+\pi^-}$, $q^2$, $\cos\theta_e$, $\cos\theta_K$, and $\chi$ are shown in
Figs.~\ref{fig:formfactor} (c-g).
In this fit, the $K^{*0}$ Breit-Wigner function follows Ref.~\cite{prl110_131802}, with a mass and width
fixed to those reported in Ref.~\cite{pdg}.  We obtain $r_V=1.67\pm0.34({\rm stat.})$ and $r_2=0.77\pm0.28({\rm stat.})$.
The fit procedure has been validated by analyzing a large inclusive MC sample,
and the pull distribution of each fitted parameter was consistent with a normal distribution.
The systematic uncertainties in the FF ratio measurements are estimated by comparing the nominal values with
those obtained after varying one source of uncertainty, as described in Ref.~\cite{prd94_032001}.
The systematic uncertainties in measuring $r_V$\,($r_2$) arise mainly from the uncertainties related to
tracking, PID and photon detection (1.8\,[2.6]\%), the $K^{*0}$ mass window (1.8\,[1.3]\%), the
${\rm MM}^2$ signal region (8.7\,[7.8]\%), the $E_{\gamma{\rm max}}$ requirement (1.2\,[1.3]\%), the
$M_{K^{*0}e^+}$ requirement (0.6\,[1.3]\%), background estimation (1.8\,[1.3]\%), and
the $K^{*0}$ Breit-Wigner line shape (0.3\,[1.3]\%).
Combining all of these in quadrature, we find the systematic uncertainties in
$r_V$ and $r_2$ of $D_s^+\rightarrow K^{*0}e^+\nu_e$ to be 9.3\% and 8.7\%, respectively.

In summary, using $3.19~\mathrm{fb}^{-1}$ data collected at $\sqrt{s}=4.178$ GeV by the BESIII detector, we measure the absolute BFs of
$D^+_s\rightarrow K^0 e^+\nu_{e}$ and $D^+_s\rightarrow K^{*0} e^+\nu_{e}$ to be
$\mathcal B({D^+_s\rightarrow K^0 e^+\nu_{e}})=(3.25\pm0.38({\rm stat.})\pm0.16({\rm syst.}))\times10^{-3}$ and
$\mathcal B({D^+_s\rightarrow K^{*0} e^+\nu_{e}})=(2.37\pm0.26({\rm stat.})\pm0.20({\rm syst.}))\times10^{-3}$.
These are the most precise measurements to date. Theoretical predictions of these BFs range from $2.0\times10^{-3}$ to
$3.9\times10^{-3}$~\cite{prd78_054002,prd71_014020,prd62_014006,ijmpa21_6125,arxiv1707,1810.11907} for
$D_s^+\rightarrow K^0e^+\nu_e$ and $1.7\times10^{-3}$ to
$2.3\times10^{-3}$~\cite{prd78_054002,prd62_014006,ijmpa21_6125,prd72_034029,arxiv1707,1810.11907} for
$D_s^+\rightarrow K^{*0}e^+\nu_e$, respectively. Since the predicated BF $2.0\times10^{-3}$ in Ref.~\cite{1810.11907} and Ref.~\cite{prd71_014020} obtained from a double-pole model are more than 2 standard deviations away from the mean value of our measured $\mathcal{B}(D_s^+\rightarrow K^0e^+\nu_e)$, thus at a confidence level of 95\%, our measurement disfavors this prediction.

By analyzing the dynamics of $D^+_s\rightarrow K^0 e^+\nu_{e}$ and $D^+_s\rightarrow K^{*0} e^+\nu_{e}$ decays
for the first time, we determine the FF of $D^+_s\rightarrow K^0 e^+\nu_{e}$ to be
$f^{K}_+(0)=0.720\pm0.084({\rm stat.})\pm0.013({\rm syst.})$ and
the FF ratios of $D^+_s\rightarrow K^{*0} e^+\nu_{e}$ to be
$r_V=1.67\pm0.34({\rm stat.})\pm0.16({\rm syst.})$ and
$r_2=0.77\pm0.28({\rm stat.})\pm0.07({\rm syst.})$.
With the FF of $D^+\rightarrow \pi^0e^+\nu_{e}$ measured by BESIII~\cite{prd96_012002} and
that of $D^+\rightarrow \rho^0 e^+\nu_{e}$ by CLEO~\cite{prl110_131802},
we calculate the ratios of the FFs of $D^+_s\rightarrow K^{0}e^+\nu_{e}$ to $D^+\rightarrow \pi^0e^+\nu_{e}$ and
$D^+_s\rightarrow K^{*0}e^+\nu_{e}$ to $D^+\rightarrow \rho^0 e^+\nu_{e}$ decays, as shown in Table~\ref{tab:ratio_ff}, which are consistent with LQCD predictions~\cite{lattice} and the expectation of $U$-spin ($d\leftrightarrow s$) symmetry~\cite{NPB492_297}.
These measurements provide a first test of the LQCD prediction that the FFs are insensitive to
spectator quarks, which has important implications when considering the corresponding $B$ and
$B_s$ decays~\cite{lattice,EPJC74_2981,prd85_114502}.

\begin{table}
\begin{center}
\caption{The ratios of the from factors.}
\resizebox{!}{1.10cm}{
\begin{tabular}
{c|c}
\hline\hline
                                                                      &  Values                                        \\
\hline
$f^{D^+_s\rightarrow K^0}_{+}(0)/f^{D^+\rightarrow\pi^0}_{+}(0)$ & $1.16\pm0.14({\rm stat.})\pm0.02({\rm syst.})$ \\
$r_V^{D^+_s\rightarrow K^{*0}}/r_V^{D^+\rightarrow\rho^0}$~~~   & $1.13\pm0.26({\rm stat.})\pm0.11({\rm syst.})$ \\
$r_2^{D^+_s\rightarrow K^{*0}}/r_2^{D^+\rightarrow\rho^0}$~~~   & $0.93\pm0.36({\rm stat.})\pm0.10({\rm syst.})$ \\
\hline \hline
\end{tabular}
}
\label{tab:ratio_ff}
\end{center}
\end{table}

The BESIII Collaboration thanks the staff of BEPCII and the IHEP computing center for their strong support. This work is supported in part by National Key Basic Research Program of China under Contract No. 2015CB856700; National Natural Science Foundation of China (NSFC) under Contracts Nos. 11335008, 11425524, 11505010, 11625523, 11635010, 11735014, 11775027; the Chinese Academy of Sciences (CAS) Large-Scale Scientific Facility Program; the CAS Center for Excellence in Particle Physics (CCEPP); Joint Large-Scale Scientific Facility Funds of the NSFC and CAS under Contracts Nos. U1532257, U1532258, U1732263; CAS Key Research Program of Frontier Sciences under Contracts Nos. QYZDJ-SSW-SLH003, QYZDJ-SSW-SLH040; 100 Talents Program of CAS; INPAC and Shanghai Key Laboratory for Particle Physics and Cosmology; German Research Foundation DFG under Contracts Nos. Collaborative Research Center CRC 1044, FOR 2359; Istituto Nazionale di Fisica Nucleare, Italy; Koninklijke Nederlandse Akademie van Wetenschappen (KNAW) under Contract No. 530-4CDP03; Ministry of Development of Turkey under Contract No. DPT2006K-120470; National Science and Technology fund; The Swedish Research Council; U. S. Department of Energy under Contracts Nos. DE-FG02-05ER41374, DE-SC-0010118, DE-SC-0010504, DE-SC-0012069; University of Groningen (RuG) and the Helmholtzzentrum fuer Schwerionenforschung GmbH (GSI), Darmstadt; This paper is also supported by the Beijing municipal government under Contract Nos. KM201610017009, 2015000020124G064, CIT\&TCD201704047, and by the Royal Society under the Newton International Fellowship Contract No. NF170002.

\end{document}